\documentclass[%
 reprint,
 amsmath,amssymb,
 aps,
prb,
]{revtex4-2}

\usepackage{graphicx}
\usepackage{dcolumn}
\usepackage{bm}
\usepackage{xspace}
\usepackage{setspace}
\usepackage{subfig}
\usepackage{verbatim}\usepackage{booktabs} 
\usepackage{siunitx}
\usepackage{amsmath}
\usepackage{ragged2e}
\usepackage{soul}
\usepackage{afterpage}

\usepackage[
    colorlinks=true,   
    linkcolor=blue,     
    citecolor=blue,     
    urlcolor=blue,      
    bookmarks=true,     
    bookmarksopen=true  
]{hyperref}

\begin{document}

\title{Superconductivity Mediated Long Range Magnetic Coupling}

\author{Mingyan Wang}
\email{wming.hqu@gmail.com} 
\affiliation{College of Information Science and Engineering, Huaqiao University, Xiamen 361021, China}

\author{Yi Liu}
\affiliation{College of Information Science and Engineering, Huaqiao University, Xiamen 361021, China}

\author{Yao Lu}
\email{yao.y.lu1203@gmail.com}
\affiliation{College of Information Science and Engineering, Huaqiao University, Xiamen 361021, China}

\date{\today}

\begin{abstract}
We investigate a Rashba superconductor thin film coupled to overlaying ferromagnetic insulators (FIs). We show that the ferromagnetic insulators generate circular super-currents, enabling long-range magnetic interactions (LRMI), decaying in power laws. In the static case, the long-range magnetic interaction can be ferromagnetic, in contrast to previous studies showing that superconductor mediates anti-ferromagnetic interactions decaying exponentially. Surprisingly, we find that in the dynamic case, the LRMI has a different distance dependence. Our results have potential applications in superconducting spintronics.

\end{abstract}

\maketitle

\section{Introduction}
Superconductors that exhibit zero electrical resistance as a result of spontaneous U(1) symmetry breaking are a central topic in condensed matter physics. Magnetic fields that align electron spins can break Cooper pairs—composed of electrons with opposite spins—and thus act as a competing order to superconductivity~\cite{anderson1959spin}. Nevertheless, superconductivity and ferromagnetism can coexist in certain systems, leading to exotic phenomena. In superconductor/ferromagnet heterostructures, Cooper pairs can penetrate the ferromagnet via the proximity effect~\cite{buzdin2005proximity}, while the ferromagnet can suppress superconductivity (the anti-proximity effect) and induce magnetization within the superconductor (magnetic proximity effect)~\cite{bergeret2004induced,xia2009inverse}. A comprehensive understanding of these proximity effects relies heavily on the quasiclassical Eilenberger and Usadel frameworks~\cite{eilenberger1968transformation,usadel1970generalized}, formulated via Gorkov equations and quasiclassical Green's functions~\cite{gor1958energy}, which have successfully predicted the generation of odd-frequency triplet Cooper pairs in the presence of inhomogeneous exchange fields. This has been extensively reviewed and experimentally verified in various hybrid structures~\cite{buzdin2005proximity,bergeret2005odd,ryazanov2001coupling,khaire2010observation,robinson2010controlled}. In noncentrosymmetric superconductors with spin–orbit coupling (SOC), a supercurrent can generate magnetization whose orientation depends on the type of SOC, a phenomenon known as the superconducting Edelstein effect~\cite{fiebig2005revival,levitov1985magnetostatics,yip2002two,fujimoto2005magnetoelectric}. In contrast, a ferromagnet in proximity to a SOC superconductor can induce a supercurrent even when the order parameter is homogeneous, giving rise to the superconducting diode effect~\cite{ando2020observation,daido2022intrinsic}.

Recently, superconducting spintronics has attracted significant attention~\cite{linder2015superconducting}. Beyond static proximity effects, the dynamic injection of spin angular momentum into superconductors—such as via ferromagnetic resonance and spin pumping—provides a powerful tool to generate pure spin currents and magnon-fluxon couplings~\cite{eschrig2015spin,jeon2018enhanced}. In F/S/F structures, the superconductor can mediate an antiferromagnetic interaction that decays exponentially over the superconducting coherence length due to the formation of triplet Cooper pairs. When one ferromagnet precesses, the other can be excited via this superconductor-mediated interaction, providing a platform to explore long-range spin control in hybrid superconducting devices. It has been shown that the equilibrium spin current carried by triplet Cooper pairs in the superconductor plays an important role.

\begin{figure}[htbp]
    \centering
    \includegraphics[width=\linewidth]{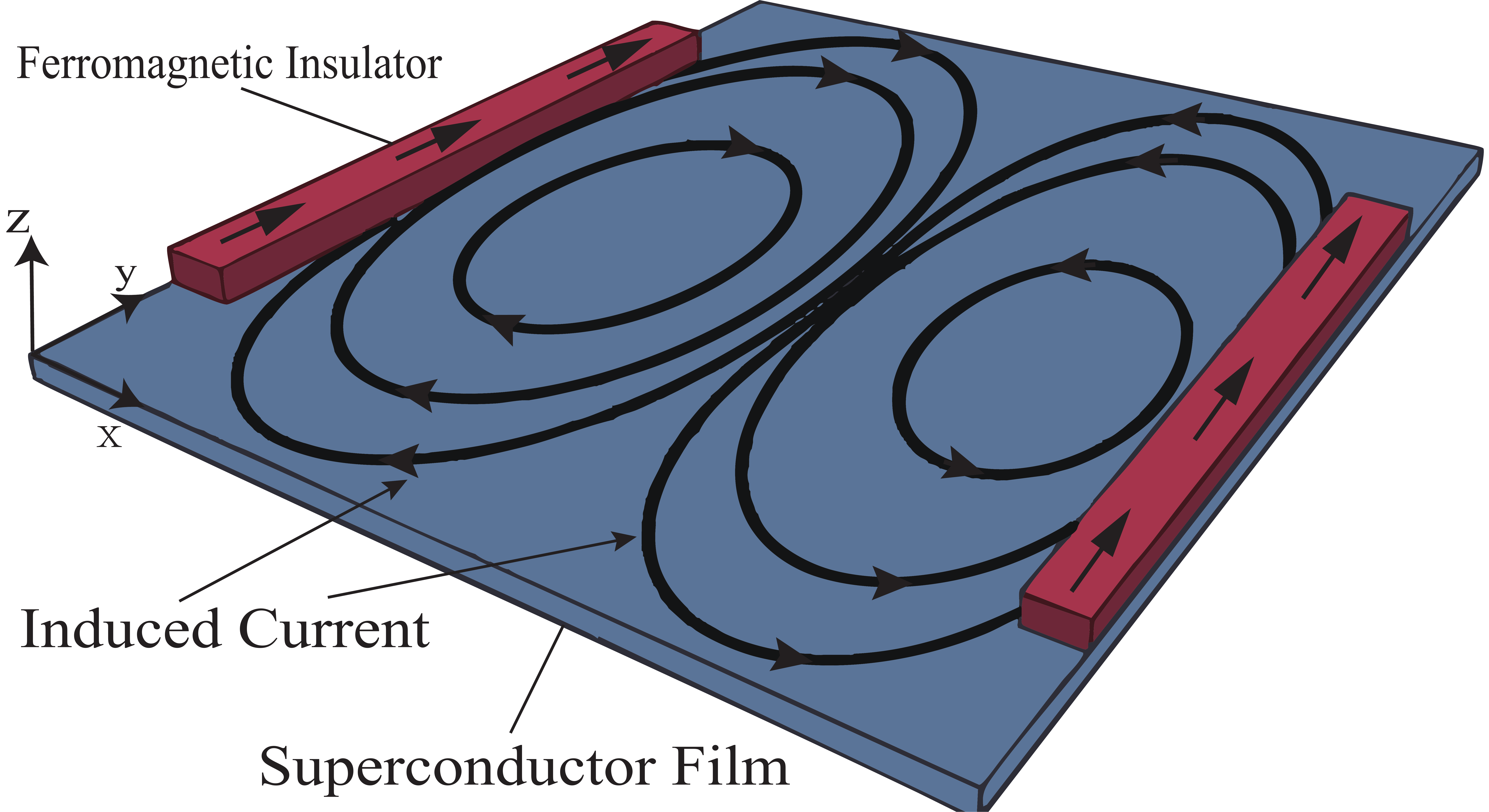}
    \caption{Schematic illustration of the induced transverse current.}
    \label{fig:j_y3D}
\end{figure}

In this work, we consider a Rashba superconductor film with two FI strips placed on the top of it, as illustrated in Fig.~\ref{fig:j_y3D}. Following the microscopic framework of magnetoelectric effects in superconductors~\cite{buzdin2008direct}, the magnetization of the FI, in combination with SOC, generates a localized anomalous current. This current alone does not satisfy the continuity equation; to restore current conservation, it induces a spatial modulation of the superconducting phase. The anomalous current, together with the supercurrent arising from the resulting phase gradient, forms closed current loops that satisfy the continuity equation. The total current exhibits power-law decay and induces magnetization in the second ferromagnet, thereby polarizing it. Consequently, the two ferromagnets interact over long distances through the superconductor. Our model differs from previous works in several key aspects: First, the LRMI can be ferromagnetic, in contrast to earlier studies where only antiferromagnetic coupling was found. Second, the LRMI is mediated by supercurrents rather than by triplet Cooper pairs. Third, the LRMI decays as a power law with distance.

This paper is organized as follows: In Section \ref{sec:model}, we introduce the theoretical model. In Section \ref{sec:static}, we analyze the static supercurrent distribution and the supercurrent-mediated LRMI between FIs. In Section \ref{sec:dynamic}, we investigate the dynamic mechanism, detailing the transverse and longitudinal components and the propagation characteristics of the far-field modes. In Section \ref{sec:conclusion}, we summarize the conclusions.

\section{Model}
\label{sec:model}

Motivated by the rapid experimental progress in isolating low-dimensional condensates~\cite{qiu2021recent}, we consider a two-dimensional (2D) superconductor film with Rashba type SOC. Two FIs are placed on top of it, located at $x=0$ and $x=r$, as depicted in Fig.~\ref{fig:j_y3D}. The superconductor film has a finite length $L$, and width $W$. The width of the FI is denoted as $w$ and the length is $l$.  We assume $w\ll l$,  and the FI can be treated as one-dimensional. Thus the magnetization of the FIs acting on the $2D$ superconductor film can be written as: $\boldsymbol{m}=\boldsymbol{m}_1\delta(x)+\boldsymbol{m}_2\delta(x-r)$, where $\boldsymbol{m}_1$ and $\boldsymbol{m}_2$ represent the magnetic moment per unit length of the respective FI strips. In the presence of SOC, the FIs generate localized anomalous currents, $\boldsymbol{j}_a=D\boldsymbol{a}$, where $D$ is the superfluid weight of the superconductor and $\boldsymbol{a}$ is the effective gauge field induced by the FIs, given by \cite{lu2023ferromagnetic}:

\begin{equation}
    \boldsymbol{a}=\Gamma J_0\boldsymbol{m}\times\hat{z},\label{a}
\end{equation}
where $\Gamma$ is a constant that depends on the SOC, and $J_0$ is the exchange interaction.

The anomalous current has two effects: first, it modifies the phase distribution of the superconductor, giving rise to circular supercurrents in the superconductor; second, it induces an electromagnetic field, which also affects the pattern of the supercurrents. The dynamics of the electromagnetic field   and phase of the superconductor are governed by the London-Maxwell equations, which in the Lorenz gauge $\nabla \cdot \boldsymbol{A} + c^{-2} \partial_t \Phi = 0$, are given by

\begin{align}
    \left( \nabla^2 - \frac{1}{c^2}\frac{\partial^2}{\partial t^2} \right) \boldsymbol{A}(\boldsymbol{r}, t) &= -\mu_0 \boldsymbol{j}(\boldsymbol{r}, t) \delta(z), \nonumber \\
    \left( \nabla^2 - \frac{1}{c^2}\frac{\partial^2}{\partial t^2} \right)\Phi (\boldsymbol{r}, t) &= -\frac{\rho(\boldsymbol{r}, t)}{\epsilon_0} \delta(z), \nonumber \\
    \boldsymbol{j} = -e^2 D &\left( \boldsymbol{A} +  \boldsymbol{a} - \frac{1}{2e} \nabla \phi \right),
    \label{eq:london}
\end{align}
where $\delta(z)$ is the Dirac delta function, $\mu_0$ is the vacuum permeability, $c$ is the speed of light, $\nabla^2$ is the Laplacian operator, and $\epsilon_0$ is the vacuum permittivity. $\boldsymbol{j}$ and $\rho$ denote the current density and charge density, respectively. $\boldsymbol{A}$ represents the vector potential and $\Phi$ stands for the electric potential respectively.

The supercurrent is modified to include the effect of the effective gauge field $\boldsymbol{a}(\boldsymbol{r}, t)$ arising from the FI via spin-orbit coupling. 
Here, $D = n_s / m^*$ is the 2D superfluid weight ($n_s$ is the superfluid density and $m^*$ is the effective mass), $e>0$ is the elementary charge, and $\phi$ is the phase of the superconductor satisfying the Josephson relation $\partial_t \phi = -2e \Phi$.

In this work, we treat the effective gauge field $\boldsymbol{a}$ as the primary source driving the system. For FI strips located at $x=0$ and $x=r$, the source term is parameterized as:
\begin{equation}
    \boldsymbol{a}(\boldsymbol{r}, t) = \boldsymbol{a}_1(t)\delta(x)+\boldsymbol{a}_2(t)\delta(x-r).
    \label{eq:source_def}
\end{equation}
By solving the London-Maxwell equations, we can obtain the supercurrent distribution in the superconductor as well as the effective magnetic interaction between the two FIs. In this paper we only consider the zero temperature case, $T=0$.

\section{Static Case}
\label{sec:static}

We first analyze the static case. In this case the magnetization  of a FI strip located at $x=0$ can be written as  $\boldsymbol{m} = (m_x \hat{x} + m_y \hat{y} + m_z \hat{z})\delta(x)$. According to Eq.~\eqref{a}, the induced effective gauge field is: $\boldsymbol{a}(\boldsymbol{r}) = (a_x \hat{x} + a_y \hat{y}) \delta(x)$, with $a_x = \Gamma J_0 m_y$, $a_y = -\Gamma J_0 m_x$. In this regime, the continuity equation $\nabla \cdot \boldsymbol{j} = 0$ imposes strict constraints on the current distribution. We first consider the short-range limit $|x| \ll r_0$, where $r_0=2/(\mu_0e^2D)$ is the Pearl length\cite{pearl1964current}, and then discuss the general case. For simplicity, we only calculate the current induced by the left FI, and the current due to the other FI has similar form.

\subsection{Short distance limit}

In the region close to the FI ($|x| \ll r_0$), the screening effect of the electromagnetic field is negligible. Thus, we can ignore the $\boldsymbol{A}$ field and the Maxwell equations. The supercurrent is governed by the continuity equation: $\nabla \cdot \boldsymbol{j} = 0$, 
and the London equation:

\begin{equation}
    \boldsymbol{j} \approx -e^2 D \left( \boldsymbol{a} - \frac{1}{2e} \nabla \phi \right).
\end{equation}

We consider a finite superconductor film of length $L$ ($x \in [-L/2, L/2]$) and width $W$ ($y \in [-W/2, W/2]$). Away from the FI ($x \neq 0$), the source field vanishes ($\boldsymbol{a}=0$), and the continuity equation implies that the phase satisfies the Laplace equation:
\begin{equation}
    \nabla^2 \phi(x,y) = 0.
\end{equation}
We assume the length of the FI is the same as the width of superconductor film $(l=W)$.
The boundary conditions are determined by the requirement that the normal component of the current vanishes at the edges ($\boldsymbol{j} \cdot \boldsymbol{n} = 0$), yielding
\begin{align}
    \frac{\partial \phi}{\partial y} \bigg|_{y=\pm W/2} &= 2e a_y \delta(x),\nonumber\\
    \frac{\partial\phi}{\partial x}\bigg|_{x=\pm L/2}&=0.
\end{align}
Solving the Laplace equation using the method of separation of variables yields the exact expressions for the current density components in the finite system:

Defining the factor $\mathcal{G}_n = \frac{\cosh(k_n y)}{\cosh(k_n W/2)}$ with $k_n = 2n\pi/L$ and $j_0 = e^2 D a_y / L$, the current distribution is expressed as:
\begin{subequations}
\begin{align}
    j_x(x,y) &= -2j_0 \sum_{n=1}^{\infty} \frac{\sinh(k_n y)}{\cosh(k_nW/2)} \sin\left(k_nx\right)\label{eq:jx_analytic}, \\
    j_y(x,y) &= -j_0 \left[L \delta(x) - \left( 1 + 2 \sum_{n=1}^{\infty} \mathcal{G}_n \cos(k_n x) \right) \right].\label{eq:jy_analytic}
\end{align}
\end{subequations}

\begin{figure}[htbp]
    \centering
    \includegraphics[width=1.05\linewidth]{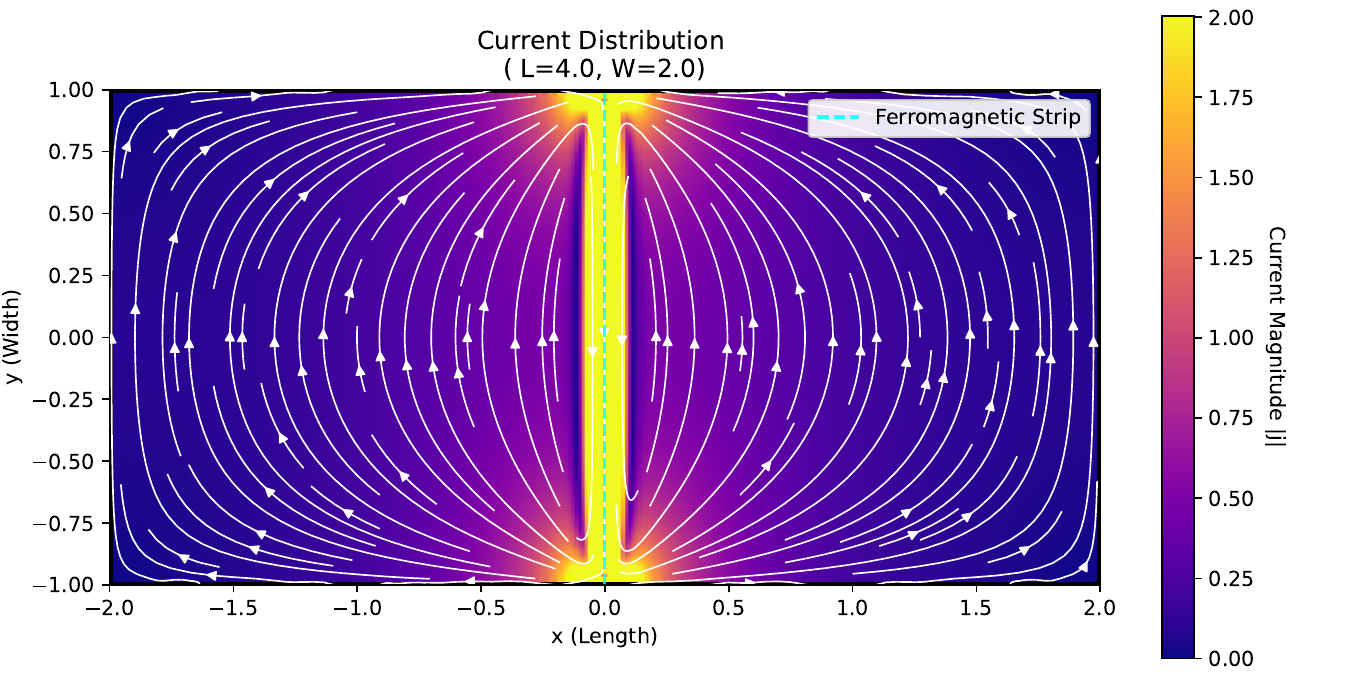}
    \caption{Streamline plot of the supercurrent distribution calculated from the analytical solution in the short distance limit Eq.~\eqref{eq:jx_analytic}. The  dashed line indicates the FI.}
    \label{fig:laplace_current}
\end{figure}

These expressions reveal the detailed structure of the current flow. The $j_y$ component Eq.~\eqref{eq:jy_analytic} contains the term $\delta(x)$ and a return current term that includes both the uniform background $1/L$ and spatially varying harmonics. The $j_x$ component Eq.~\eqref{eq:jx_analytic} arises from the phase gradient required to satisfy the boundary conditions.
The result is illustrated in Fig.~\ref{fig:laplace_current}. 

To generalize these findings, we consider a FI located at an arbitrary coordinate $x = d$ within the domain defined by $x \in [0, L]$ and $y \in [-W/2, W/2]$, where $W$ and $L$ are the dimensions of the superconductor film along the $y$- and $x$-directions, respectively.
To express solutions more compactly, we introduce $C_m = \cos(q_m d) / \cosh(q_m W/2)$. The induced currents $\boldsymbol{j}^{\text{res}}$ then simplify to:
\begin{subequations}
\begin{align}
j_x^{\text{res}}(x,y) &= -2j_0 \sum_{m=1}^{\infty} C_m \sinh(q_m y) \sin(q_m x),\label{j_x} \\
j_y^{\text{res}}(x,y) &= j_0 \left[ 1 + 2 \sum_{m=1}^{\infty} C_m \cosh(q_m y) \cos(q_m x) \right],\label{j_y}
\end{align}
\end{subequations}
where $q_m = m\pi/L$. The total current is given by $\boldsymbol{j} = \boldsymbol{j}^{\text{s}} + \boldsymbol{j}^{\text{res}}$, where $j_y^{\text{s}} = -e^2 D a_y \delta(x-d)$ and $j_x^{\text{s}} = 0$. This solution naturally recovers the symmetric case when the FI is centered at $d = L/2$.

\begin{figure}[htbp]
    \centering
    \includegraphics[width=1.1\linewidth]{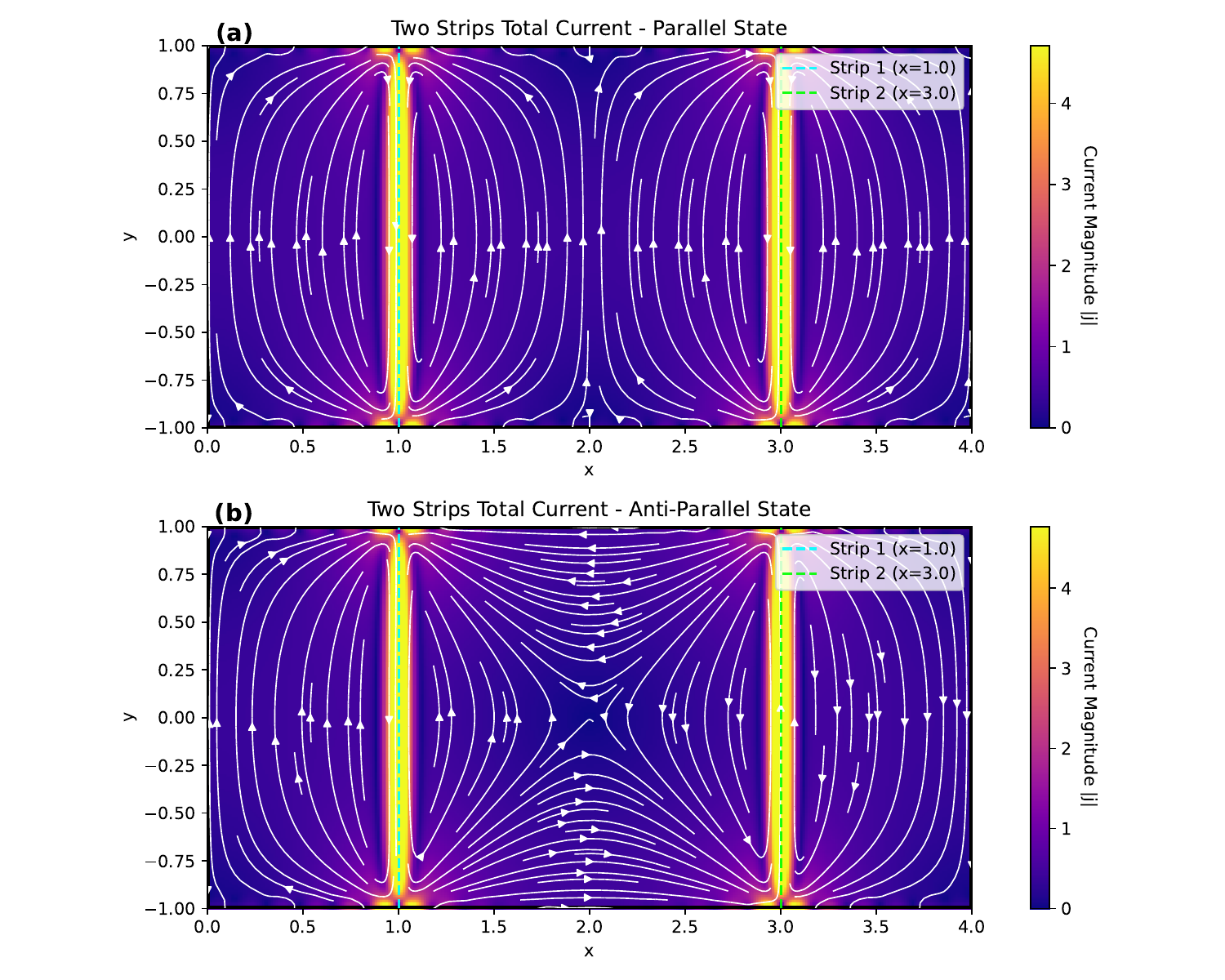}
    \caption{Spatial distribution of the total induced current $\boldsymbol{j}(x,y)$ driven by two FIs located at $d_1=1.0$ and $d_2=3.0$ with $L=4,W=2$. (a) Parallel state. (b) Anti-parallel state.}
    \label{fig:two_strips}
\end{figure}

The  supercurrent distribution in the superconductor in the presence of two FIs is shown in Fig.~\ref{fig:two_strips}. One can see that they interact with each other via the supercurrents.

\subsection{General Case \& Long Distance Current}

At distances comparable to or exceeding the Pearl length, we have to take into account the screening effect of electromagnetic field $\boldsymbol{A}$ induced by the supercurrents. For simplicity, we assume the size of the superconductor is much larger than $r_0$, so that we can treat it as an infinite plane. In this case, we need to solve the full Maxwell-London equation Eq.~\eqref{eq:london}. The solution of Eq.~\eqref{eq:london} in Fourier space is given by:
\begin{align}
    {j}_y(q_x) = & -e^2 D a_y \frac{r_0 |q_x|}{1 + r_0 |q_x|}\nonumber\\
    {j}_x(q_x) = & 0.
    \label{eq:jy_momentum}
\end{align}
The detailed derivation of Eq.~\eqref{eq:jy_momentum} is provided in Appendix \ref{eq:9}. 
The real-space current distribution is obtained by performing the inverse Fourier transform:
\begin{equation}
    j_y(x) = -e^2 D a_y \int_{-\infty}^{\infty} \frac{dq_x}{2\pi} e^{iq_xx} \left( 1 - \frac{1}{1 + r_0 |q_x|} \right).
    \label{eq:jy_integral}
\end{equation}
Evaluating this integral yields the form:
\begin{equation}
    j_y(x) = -e^2 D a_y \left[ \delta(x) - \frac{1}{\pi r_0} \mathcal{F}\left(\frac{|x|}{r_0}\right) \right],
    \label{eq:jy_analytical}
\end{equation}
where:
\begin{equation}
    \mathcal{F}(x) = \sin(x)[\frac{\pi}{2} - \operatorname{Si}(x)] - \cos(x)\operatorname{Ci}(x)\label{mathcalF},
\end{equation}

The cosine integral is defined as $\operatorname{Ci}(x)=-\int_{x}^{\infty}\frac{\cos t}{t}\mathrm{d}t$ and the sine integral as $\operatorname{Si}(x)=\int_{0}^{x}\frac{\sin t}{t}\mathrm{d}t$.

In the far-field limit ($|x| \gg r_0$), the asymptotic expansion of the structure function reveals a power-law decay:
\begin{equation}
    j_y(x) \approx \frac{e^2 D a_y}{\pi r_0} \left( \frac{r_0}{x} \right)^2 \propto \frac{1}{x^2}.
    \label{eq:jy_asymptotic}
\end{equation}

This slow $1/x^2$ decay is the hallmark of the 1D cross-section of a Pearl vortex. 

\subsection{Interaction}

To describe the interaction between the two FI strips, we start with the total effective interaction Hamiltonian for a two-dimensional superconductor with SOC and exchange field  \cite{lu2023ferromagnetic}:
\begin{equation}
H_{eff} = \int d^{3}r \left\{ \frac{1}{8}D [\nabla \phi - 2e\boldsymbol{A} - 2e\boldsymbol{a}]^2 \delta(z) + \frac{1}{2\mu_0} \boldsymbol{B}^2 \right\}, \label{eq:F_total}
\end{equation}
the expression simplifies to a form determined by the supercurrent $\boldsymbol{j}$ in the region where the exchange field is finite \cite{lu2023ferromagnetic}:
\begin{equation}
H_{eff} = -\frac{1}{2} \int d^2\boldsymbol{r} \, \boldsymbol{j} \cdot \boldsymbol{a} \label{eq:F_simplified}.
\end{equation}
Here, $\boldsymbol{j}=\boldsymbol{j}_1+\boldsymbol{j}_2$ and $\boldsymbol{a}=\boldsymbol{a}_1+\boldsymbol{a}_2$. Where $\boldsymbol{j}_1$ is the supercurrent induced by the left FI and $\boldsymbol{j}_2$ for the right FI. $\boldsymbol{a}_1$ is the effective gauge field induced by the left FI and $\boldsymbol{a}_2$ for the right FI. We consider the system is infinitely long in the $x$-direction and sufficiently large in the $y$-direction. Thus, the interaction depends only on the distance($r$) between the two FIs. We define the interaction energy $H_{eff}(r)$ as the cross term of Eq.~\eqref{eq:F_simplified}. Here we use $\boldsymbol{j}_1(x_2)\cdot\boldsymbol{a}_2=\boldsymbol{j}_2(x_1)\cdot\boldsymbol{a}_1$:
\begin{equation}
    H_{eff}(r) = -\iint dx dy\, \boldsymbol{j}_1(x) \cdot \boldsymbol{a}_2(x)\label{E},
\end{equation}
Substituting Eq.~\eqref{eq:jy_analytical} into Eq.~\eqref{E} yields the result:
\begin{equation}
    H_{eff}(r) = -\frac{J_e  M_{1x}M_{2x}}{l r_0} \mathcal{F}\left(\frac{r}{r_0}\right),
\end{equation}

where $\mathcal{F}(x) $ is defined by Eq.~\eqref{mathcalF}, and $J_e=e^2\Gamma^2J_0^2D/\pi$,  $M_{1x}=lm_{1x}(M_{2x}=lm_{2x})$ is the total magnetic moment of FI 1 (FI 2), $r$ is the distance between two FIs, $l$ is the  length of the FI. The relationship between $H_{eff}(r)$ and distance is determined  by $H_{eff}(r) \propto\mathcal{F}(r/r_0)$. We have analyzed its far-field behavior in Eq.~\eqref{eq:jy_asymptotic}. When $x$ is very small,  we have $\operatorname{Si}(x) \approx x - \frac{x^3}{18} + \cdots \approx x$, $\operatorname{Ci}(x) \approx \gamma + \ln(x) - \frac{x^2}{4} + \cdots \approx \gamma + \ln(x)$, $\sin(x) \approx x$, $\cos(x) \approx 1$, where $\gamma$ is the Euler–Mascheroni constant. So we can write  $\mathcal{F}(x)$ when $x$ is very small as:
$\mathcal{F}(x) \approx -\ln(x) - \gamma + \frac{\pi}{2}x$. Thus, we can find the short-range interaction ($r \ll r_0$) exhibits a logarithmic divergence, $F(r) \propto \ln(r/r_0)$. In the far-field limit ($r \gg r_0$), the interaction transitions to  power-law decay, $F(r) \propto (r_0/r)^2$. Unlike previous studies, our interaction is ferromagnetic rather than antiferromagnetic, and the interaction exhibits $x^{-2}$ power-law decay at large distances.

\section{Dynamic Situation}
\label{sec:dynamic}

In this section, we consider the case in which the FI is precessing. We assume that the easy axis of the FIs is $z$-axis. The ground state of the system is that the magnetizations of both FIs are parallel to the $z$-axis. Upon excitation, the magnetization precesses and acquires time-dependent oscillating components $m_x$ and $m_y$ in the  $xy$ plane. Consequently, the time-dependent magnetization can be expressed as $\boldsymbol{m}(t) = [m_z \hat{z} + (m_x \hat{x} + i m_y \hat{y})e^{-i\omega t}]\delta(x)$.

According to the relation $\boldsymbol{a}=\Gamma J_0\boldsymbol{m}\times\hat{z}$, these dynamic magnetization  components induce a  time-dependent effective gauge field. The effective  gauge field generated by the FI is therefore given by $\boldsymbol{a}(\boldsymbol{r},t) = (a_{x}\hat{x} + ia_{y}\hat{y})\delta(x)e^{-i\omega t}$, where $a_x = i \Gamma J_0 m_y$, $ a_y = i \Gamma J_0 m_x$. We assume the size of the superconductor is much larger than $r_0$, so that we can treat it
as an infinite plane.

Unlike the static case in which the supercurrent has only transverse component, in the dynamic case the time dependent supercurrent has both longitudinal and transverse components. The supercurrent can be written as $\boldsymbol{j}=\boldsymbol{j}_l+\boldsymbol{j}_t$, where $\boldsymbol{j}_t$ is the transverse component of $\boldsymbol{j}$ satisfying $ \nabla \cdot \boldsymbol{j}_t = 0$ and $\boldsymbol{j}_l$ is the longitudinal component of $\boldsymbol{j}$ satisfying $ \nabla \times \boldsymbol{j}_l = 0$.
In the following, we analyze the transverse component and longitudinal component of the supercurrent to establish the rules for long-range coupling.

\subsection{Transverse component}
We perform a two dimensional Fourier transform for $\boldsymbol{A},\boldsymbol{a},\boldsymbol{j},\Phi,\phi$ by introducing $\boldsymbol{q} = (q_x,q_y)$. Since the system is considered homogeneous in the $y$-direction, only the $q_y=0$ component is non-zero after performing a Fourier transformation. So we can write $\boldsymbol{q}$ as $\boldsymbol{q}=(q_x,0)$. The transverse current $\boldsymbol{j}_t=j_y\hat{y}$ is driven by the component $a_y$. The first equation of Eq.~\eqref{eq:london} in the $y/x$-direction can be written in Fourier space as:

\begin{equation}
    \left( \frac{\partial^2}{\partial z^2} - \kappa^2 \right) A_{y/x}(q_x, z) = -\mu_0 j_{y/x}(q_x) \delta(z),\label{1}
\end{equation}
where $\kappa(q_x) = \sqrt{q_x^2 - k_0^2}$, $k_0=\omega/c$. The solution to Eq.~\eqref{1} at $z \neq 0$ is in exponentially decaying form: $A_{y/x}(q_x, z) = A_{y/x}(q_x, 0) e^{-\kappa|z|}$. For the sake of brevity, we will shorten $A_{y/x}(q_x, 0)$ to $A_{y/x}(q_x)$.

The current $\boldsymbol{j}$ in Fourier space is:

\begin{align}
    \boldsymbol{j}(q_x) = -e^2 D \left[ \boldsymbol{A}(q_x) + \boldsymbol{a}(q_x) - \frac{i\boldsymbol{q}}{2e}\phi(q_x) \right]\label{2}.
\end{align}

We find the relationship between $A_{y/x}$ and $j_{y/x}$ is $ 2\kappa A_{y/x}(q_x) = \mu_0 j_{y/x}(q_x)$. Then we obtain:
\begin{align}
    j_y(q_x) &= -e^2 D \left[ \frac{\mu_0}{2\kappa} j_y(q_x) + i a_y \right],\\
    j_x(q_x) &= -e^2 D \left[ A_x(q_x) + a_x - \frac{iq_x}{2e}\phi(q_x) \right]
\end{align}
where $\kappa(q_x) = \sqrt{q_x^2 - k_0^2}$, $k_0=\omega/c$.
Its integral representation in real space is given by:
\begin{equation}
    j_y(x, t) = e^2 Di a_y e^{-i\omega t} \int_{-\infty}^{\infty} \frac{dq_x}{2\pi} e^{iq_xx} \left[\frac{1}{1 + r_0 \kappa(q_x)}-1\right],
    \label{eq:jy_integral2}
\end{equation}
 The detailed derivation of Eq.~\eqref{eq:jy_integral2} is provided in Appendix \ref{sec:j_xj_y}. Since the integral $\int_{-\infty}^{\infty} \frac{dq_x}{2\pi} e^{iq_xx}1/(1 + r_0 \kappa(q_x))$ is conditionally convergent, we need to introduce $\omega \rightarrow \omega +i0^+$ to ensure the causality. The first-order term asymptotically extended in the far field($k_0 x \gg 1$) can be obtained as follows:

\begin{equation}
    j_y(x, t) \approx - \left[ \frac{ie^2 D a_y r_0 \sqrt{k_0}}{\sqrt{2\pi}} \right] \frac{e^{i(k_0 |x| - \omega t + 3\pi/4)}}{|x|^{3/2}},
    \label{eq:jy_t_asymptotic}
\end{equation}

Note that this decay is slower than the $x^{-2}$ static case derived previously, indicating that dynamic current can propagate over longer distances. This term vanishes in the zero-frequency limit $\omega \to 0$. As $\omega \to 0$, the radiative contribution disappears, and the system seamlessly recovers the static solution ($j_y \propto x^{-2}$) described by Eq.~\eqref{eq:jy_asymptotic}.

\subsection{Longitudinal Component}

The oscillation of the longitudinal component of the supercurrent leads to charge accumulation. By solving Eq.~\eqref{eq:london} using Eq.~\eqref{2} and the Josephson relation $\partial_t \phi = -2e \Phi$ to eliminate $\Phi$, we can obtain the relation between $\phi$ and $A_x$ :
\begin{equation}
    \phi(q_x) = \frac{2e c^2 q_x}{i \omega^2} A_x(q_x).
    \label{22}
\end{equation}

By combining $2\kappa A_x(q_x) = \mu_0 j_x(q_x)$ with Eq.~\eqref{2} and Eq.~\eqref{22}, we obtain the longitudinal supercurrent $\boldsymbol{j}_l=j_x\hat{x}$ in Fourier space as:

\begin{equation}
    {j}_x(q_x) = e^2 D \frac{r_0 \omega^2}{c^2\kappa - r_0\omega^2} a_x,
    \label{eq:longitudinal_j}
\end{equation}
where $\kappa(q_x) = \sqrt{q_x^2 - k_0^2}$, $k_0=\omega/c$. The detailed derivation of Eq.~\eqref{eq:longitudinal_j} is provided in Appendix \ref{sec:j_xj_y}.  The poles of Eq.~\eqref{eq:longitudinal_j} are determined by $c^2\kappa - r_0\omega^2=0$, leading to:
\begin{equation}
q_x^2 = \frac{\omega^2}{c^2} + r_0^2 \frac{\omega^4}{c^4}
\end{equation}
The asymptotic expansion in the far field ($k_0|x| \gg 1$) can be written as:
\begin{align}
    j_x(x,t) &\approx   i e^2 D a_x  \frac{\alpha^2}{ q_p}  e^{i(q_p |x| - \omega t)}\nonumber\\
    -&\left[ \frac{e^2 D a_x \sqrt{k_0}}{\sqrt{2\pi} \alpha} \right] \frac{e^{i(k_0|x| - \omega t + 3\pi/4)}}{|x|^{3/2}}\label{j_xt},
\end{align}
where $q_p = \sqrt{k_0^2 + \alpha^2}$, $k_0=\omega/c$, $\alpha= r_0 \omega^2/c^2$.
Eq.~\eqref{j_xt} demonstrates that the longitudinal current consists of two components. The first term originates from the pole condition $c^2\kappa - r_0\omega^2 = 0$. The second term exhibits the same $|x|^{-3/2}$ spatial decay as the transverse component. We find current flowing in the x-direction under dynamic case, which is not present under static case. 

\subsection{Interaction In the Dynamic Case}
In the dynamic regime, we consider two FIs, denoted as FI 1 and FI 2, placed on the top of the superconductor film. FI 1 is located at $x=0$, and FI 2 is located at $x=r$. The magnetization generated by FI 1 is: $\boldsymbol{m}_1(t) =[m_{1z} \hat{z} + (m_{1x} \hat{x} + i m_{1y} \hat{y})e^{-i\omega_1 t}]\delta(x)$, and the magnetization generated by FI 2 is: $\boldsymbol{m}_2(t) =[ m_{2z} \hat{z} + (m_{2x} \hat{x} + i m_{2y} \hat{y})e^{-i\omega_2 t}]\delta(x-r)$ . According to Eq.~\eqref{a} in our model, the  effective gauge field, which is induced by FI 1 is given by $\boldsymbol{a}_{1}(\boldsymbol{r}, t) = (a_{1x}\hat{x} + i a_{1y}\hat{y})\delta(x)e^{-i\omega_1 t}$, with components $a_{1x} = i \Gamma J_0 m_{1y}$ and $a_{1y} = i \Gamma J_0 m_{1x}$, and the effective gauge field for FI 2 is $\boldsymbol{a}_{2}(\boldsymbol{r}, t) = (a_{2x}\hat{x} + i a_{2y}\hat{y})\delta(x-r)e^{-i\omega_2 t}$, with components $a_{2x} = i \Gamma J_0 m_{2y}$ and $a_{2y} = i \Gamma J_0 m_{2x}$. We can introduce the effective interaction Hamiltonian $H_{eff}(r,t)$ as $H_{eff}(r,t)=F_{int}(t)$. By substituting the derived dynamic supercurrent  into the Eq.~\eqref{E}, we obtain:
\begin{align}
    &H_{eff}  = -\iint dx dy\, \boldsymbol{j}_1\cdot\boldsymbol{a}_2
    =- \frac{l(\Gamma J_0)^2}{2} \operatorname{Re} \int_{-\infty}^{\infty} \frac{dq_x}{2\pi} e^{iq_x r} \nonumber\\
    & \left\{ \left[ m_{1y} m_{2y} K_x(q_x, \omega_1) - i m_{1x} m_{2x} K_y(q_x, \omega_1) \right] e^{-i(\omega_1 - \omega_2)t} \right. \nonumber\\
    & \left. - \left[ m_{1y} m_{2y} K_x(q_x, \omega_1) + i m_{1x} m_{2x} K_y(q_x, \omega_1) \right] e^{-i(\omega_1 + \omega_2)t} \right\}.\label{int_dy}
\end{align}
Here, the kernels at the driving frequency $\omega_1$ are $K_x(q_x, \omega_1) = e^2 D \frac{r_0 \omega_1^2}{c^2\kappa_1 - r_0\omega_1^2}$ and $K_y(q_x, \omega_1) = -i e^2 D \frac{r_0 \kappa_1}{1 + r_0 \kappa_1}$, where $\kappa_1 = \sqrt{q_x^2 - \omega_1^2/c^2}$.

Due to the effective interaction Hamiltonian $H_{eff}$, a magnon excited in one FI can induce the excitation of a magnon in the other FI.

Furthermore, in Appendices \ref{sec:time_dependent}-\ref{Far-field Asymptotic Expansion}, we present a brief extension of the ferromagnetic impurity case from \cite{lu2023ferromagnetic} by incorporating time dependence, and we also consider the scenario of a spherical superconducting thin film.

\section{Conclusion}
\label{sec:conclusion}

In summary, this paper investigates a two-dimensional superconductor film with Rashba spin-orbit coupling interacting with FIs. The combination of magnetization and spin-orbit coupling generates localized anomalous currents. In the static regime, we consider the film is infinitely large, we demonstrate these supercurrents mediate a long-range ferromagnetic interaction between the FIs. This interaction decays as a power law with distance, specifically following a $1/x^{2}$ dependence for the static case. In the dynamic regime for FIs on an infinite film, magnons excited in one insulator are transported to another. The dynamic transverse current exhibits a $|x|^{-3/2}$ spatial decay, demonstrating that dynamic modes propagate over longer distances than static modes. 

\begin{acknowledgments}

     Y. L. acknowledges the support from Huaqiao University (605-50Y24031) and the support from Natural Science Foundation of Xiamen China (605-52424128).
     \end{acknowledgments}

\bibliographystyle{apsrev4-2}

\bibliography{manuscript}


\appendix
\onecolumngrid
\renewcommand{\theequation}{\thesection\arabic{equation}}
\renewcommand{\thefigure}{S\arabic{figure}}
\renewcommand{\thetable}{S\arabic{table}}
\begin{center}
    \textbf{\large }\\[0.3cm]
    
\end{center}

\vspace{0.8cm}

\section{Derivation of current in Fourier space for static case}
\label{eq:9}
In the static limit, the supercurrent density is governed by the relation \cite{lu2023ferromagnetic}:
\begin{equation}
    \boldsymbol{j} = -e^2 D(\boldsymbol{A} + \boldsymbol{a}_t)\label{jjjjj}.
\end{equation}

The vector potential $\boldsymbol{A}$ satisfies the Maxwell equation $\nabla^2 \boldsymbol{A} = -\mu_0 \boldsymbol{j}(x,y) \delta(z)$. 

Applying a two-dimensional Fourier transform in the $xy$-plane, the equation for the Fourier component $\boldsymbol{A}(\boldsymbol{q}, z)$ becomes:
\begin{equation}
    \left( \frac{\partial^2}{\partial z^2} - \boldsymbol{q}^2 \right) \boldsymbol{A}(\boldsymbol{q}, z) = -\mu_0 \boldsymbol{j}(\boldsymbol{q}) \delta(z)
\end{equation}
where $\boldsymbol{q}^2 = q_x^2 + q_y^2$. Since the system is considered homogeneous in the $y$-direction, we can write $\boldsymbol{q}$ as $\boldsymbol{q}=(q_x,0)$. The solution that decays as $|z| \to \infty$ is given by $\boldsymbol{A}(q_x, z) = \boldsymbol{A}(q_x, 0) e^{-|q_x||z|}$. Integrating the differential equation across the $z=0$ plane yields the condition for the derivative:
\begin{equation}
    \left. \frac{\partial \boldsymbol{A}}{\partial z} \right|_{z=0^+} - \left. \frac{\partial \boldsymbol{A}}{\partial z} \right|_{z=0^-} = -\mu_0 \boldsymbol{j}(q_x)
\end{equation}
Substituting the exponential solution leads to $-2|q_x| \boldsymbol{A}(q_x, 0) = -\mu_0 \boldsymbol{j}(q_x)$, which establishes the relation $\boldsymbol{A}(q_x) = \frac{\mu_0}{2|q_x|} \boldsymbol{j}(q_x)$. 

Inserting this expression for $\boldsymbol{A}(q_x)$ back into the constitutive relation for the $y$-component, $j_y(q_x) = -e^2 D (A_y(q_x) + a_y)$, we obtain:
\begin{equation}
    j_y(q_x) = -e^2 D \left( \frac{\mu_0}{2|q_x|} j_y(q_x) + a_y \right)
\end{equation}
Rearranging the terms to solve for $j_y(q_x)$ yields:
\begin{equation}
    j_y(q_x) \left( 1 + \frac{\mu_0 e^2 D}{2|q_x|} \right) = -e^2 D a_y
\end{equation}
By introducing the Pearl length  $r_0 = 2 / (\mu_0 e^2 D)$, the expression simplifies to:
\begin{equation}
    j_y(q_x) \left( 1 + \frac{1}{r_0 |q_x|} \right) = -e^2 D a_y \implies j_y(q_x) = -e^2 D a_y \frac{r_0 |q_x|}{1 + r_0 |q_x|}
\end{equation}
This confirms the result presented in  Eq.~\eqref{eq:jy_momentum}. For the longitudinal component, since the transverse source $a_{tx} = 0$, it follows that $j_x(q_x) = 0$.
\section{Derivation of the current in sufficiently large superconductor film}
\label{sec:j_xj_y}

We start our derivation from Eq.~\eqref{eq:london}. We assume that all quantities that change with time have a simple harmonic form $e^{-i\omega t}$. As the source term is located $x=0$, so we can obtain the result of $\boldsymbol{a}(\boldsymbol{r})$ using Fourier transform:$\tilde{\boldsymbol{a}}(q) = a_x \hat{x} + i a_y \hat{y}$. We first solve for the current distribution in the y-direction. Taking a Fourier transform of the equation of the vector potential $\boldsymbol{A}$ yields:
\begin{equation}
    \left( \frac{\partial^2}{\partial z^2} - \kappa^2 \right) A_y(q_x, z) = -\mu_0 j_y(q_x) \delta(z)\label{A_z}
\end{equation}
where $\kappa^2 = q_x^2-k_0^2$, $k_0=\omega/c$. 

The solution to Eq.~\eqref{A_z} at  $z \neq 0$ is in exponentially decaying form: $A_y(q_x, z) = A_y(q_x, 0) e^{-\kappa|z|}$. For the sake of brevity, we will shorten $A_y(q_x, 0)$ to $A_y(q_x)$.
The current $\boldsymbol{j} $ in Fourier space is of the form:
\begin{equation}
    \boldsymbol{j}(q_x) = -e^2 D \left[ \boldsymbol{A}(q_x) + \boldsymbol{a}(q_x) - \frac{i\boldsymbol{q}}{2e}\phi(q_x) \right],
\end{equation}
where $\boldsymbol{q}=(q_x,q_y)$, $q_y = 0$, so we find:
\begin{align}
    j_y(q_x) &= -e^2 D \left[ A_y(q_x) + i a_y \right]\label{j_yq}\\
    j_x(q_x) &= -e^2 D \left[ A_x(q_x) + a_x - \frac{iq_x}{2e} \phi(q_x) \right]
\end{align}

Integrating both sides of Eq.~\eqref{A_z} yields:
\begin{equation}
    \frac{\partial A_y}{\partial z}\Big|_{0^+} - \frac{\partial A_y}{\partial z}\Big|_{0^-} = -\mu_0 j_y(q_x)\label{A_relati}
\end{equation}

Substituting  $A_y(q_x, z) = A_y(q_x) e^{-\kappa|z|}$ into Eq.~\eqref{A_relati}, we get: $ 2\kappa A_y(q_x) = \mu_0 j_y(q_x)$. The calculation in the x-direction can be obtained similarly. Then we substitute this relationship into Eq.~\eqref{j_yq}: $j_y(q_x) = -e^2 D \left[ A_y(q_x) + i a_y \right]$ and we get:
\begin{equation}
    j_y(q_x) = -e^2 D \left[ \frac{\mu_0}{2\kappa} j_y(q_x) + i a_y \right].
\end{equation}
Solving this will yield:
\begin{equation}
    {j}_y(q_x) = -i e^2 D a_y \frac{r_0 \kappa(q_x)}{1 + r_0 \kappa(q_x)},
\end{equation}
where $r_0=2/(\mu_0e^2D)$ is the Pearl length. Convert it to real space:
\begin{align}
    j_y(x,t) &= -i e^2 D a_y e^{-i\omega t} \int_{-\infty}^{\infty} \frac{dq_x}{2\pi} e^{iq_xx} \frac{r_0 \sqrt{q_x^2 - k_0^2}}{1 + r_0 \sqrt{q_x^2 - k_0^2}}.\\
    & = -i e^2 D a_y e^{-i\omega t} \left[ \delta(x) - \int_{-\infty}^{\infty} \frac{dq_x}{2\pi} e^{iq_xx} \frac{1}{1 + r_0 \sqrt{q_x^2 - k_0^2}} \right]
    \label{j_yt}
\end{align}

Next, we will discuss the current distribution in the x-direction. Write our Lorenz gauge conditions in Fourier space:
\begin{equation}
    iq_x A_x(q_x) - i\frac{\omega}{c^2} \Phi(q_x) = 0 \implies \Phi(q_x) = \frac{c^2 q_x}{\omega} A_x(q_x),\label{phi}
\end{equation}
It should be noted that since the $y$-direction is infinite, the term $\partial A_y/\partial y $ becomes 0. Josephson relations in Fourier space are:
\begin{equation}
    -i\omega \phi(q_x) = -2e \Phi(q_x) \implies \phi(q_x) = \frac{2e}{i\omega} \Phi(q_x)\label{jose}
\end{equation}

By combining Eq.~\eqref{jose} and Eq.~\eqref{phi}, we can obtain:
\begin{equation}
    \phi(q_x) = \frac{2e c^2 q_x}{i \omega^2} A_x(q_x)
\end{equation}

We write $j_x$ in Fourier space:
\begin{align}
    j_x(q_x) &= -e^2 D \left[ A_x(q_x) + a_x - \frac{iq_x}{2e} \phi(q_x) \right]\nonumber\\
    &= -e^2 D \left[ A_x(q_x) + a_x - \frac{iq_x}{2e} \left( \frac{2e c^2 q_x}{i \omega^2} A_x(q_x) \right) \right]\nonumber\\
    &= -e^2 D \left[ A_x(q_x) \left( 1 - \frac{c^2 q_x^2}{\omega^2} \right) + a_x \right]\nonumber\\
    & = -e^2 D \left[ -\frac{c^2 \kappa^2}{\omega^2} A_x(q_x) + a_x \right]
\end{align}

Eq.~\eqref{A_relati} also applies to the x-direction, so we can obtain: $ 2\kappa A_x(q_x) = \mu_0 j_x(q_x)$. Therefore, $j_x(q_x)$ can be simplified to:
\begin{equation}
    j_x(q_x) = -e^2 D \left[ -\frac{c^2 \kappa^2}{\omega^2} \left( \frac{1}{r_0 e^2 D \kappa} j_x(q_x) \right) + a_x \right]\label{j_x(q)}
\end{equation}
where $r_0$ is the Pearl length. We can obviously solve for the relationship between $j_x(q_x)$ and $a_x$ using the Eq.~\eqref{j_x(q)}:
\begin{equation}
    {j}_x(q_x) = e^2 D \frac{r_0 \omega^2}{c^2 \kappa - r_0 \omega^2} a_x\label{j_xq}
\end{equation}

\section{Asymptotic derivation of far-field current of \texorpdfstring{$\boldsymbol{j}_y$}{j\_y}}

The current distribution $\boldsymbol{j}$ is given by Eq.~\eqref{j_xq} and Eq.~\eqref{j_yt}. For $x>0$, the Dirac delta function $\delta(x)= 0$, so we just need to calculate the integral in the far-field limit $x\gg0$.
\begin{equation}
    I=\int_{-\infty}^{\infty} \frac{\mathrm{d}q_x}{2\pi} e^{iq_xx} \frac{1}{1 + r_0 \sqrt{q_x^2 - k_0^2}}
    \label{eq:I_origin}
\end{equation}
For the sake of causality, we introduce $\omega \rightarrow \omega +i0^+$, which implies $k_0 \rightarrow k_0 +i0^+$. 

To evaluate this integral, we deform the integration contour from the real axis into the upper half-plane (UHP). Since the integrand vanishes as $|q_x| \to \infty$, Jordan's lemma applies safely, and the integration along the real axis can be replaced by the integral along the ``hairpin'' contour $\Gamma_{\mathrm{cut}}$ wrapping around the branch cut (i.e., $I_{\mathrm{real}} = -I_{\mathrm{cut}}$). In the far-field limit ($x \gg 0$), the exponential decay $e^{-sx}$ along the cut strictly confines the dominant contribution to the immediate vicinity of the branch point $q_x \approx k_0$.

Within this local region, we can now expand the denominator using the Taylor series $\frac{1}{1+x}=1-x+x^2-\dots$:
\begin{equation}
      \frac{1}{1 + r_0 \sqrt{q_x^2 - k_0^2}} = 1 - r_0 \sqrt{q_x^2 - k_0^2} + r_0^2 (q_x^2 - k_0^2) - \dots
\end{equation}
The constant term $1$ is continuous across the cut and contributes nothing to the branch cut integral (or equivalently, yields a $\delta(x)$ which is $0$ for $x > 0$). Thus, the long-range asymptotic behavior is dominated by the leading non-trivial term evaluated along $\Gamma_{\mathrm{cut}}$. Eq.~\eqref{eq:I_origin} becomes:
\begin{equation}
    I \approx -\int_{\Gamma_{\mathrm{cut}}} \frac{\mathrm{d}q_x}{2\pi} e^{iq_xx} \left( -r_0 \sqrt{q_x^2 - k_0^2} \right) = -\frac{r_0}{2\pi} \left( -\int_{\Gamma_{\mathrm{cut}}} e^{iq_xx} \sqrt{q_x^2 - k_0^2} \mathrm{d}q_x \right)
    \label{eq:I_approx}
\end{equation}

Since the dominant contribution arises from $q_x \approx k_0$, we can approximate the square root term as:
\begin{equation}
    \sqrt{q_x^2 - k_0^2} = \sqrt{q_x - k_0}\sqrt{q_x + k_0} \approx \sqrt{2k_0} \sqrt{q_x - k_0}
\end{equation}
Substituting this approximation back into Eq.~\eqref{eq:I_approx}, we extract the constant prefactor:
\begin{equation}
    I \approx -\frac{r_0 \sqrt{2k_0}}{2\pi} \left( -\int_{\Gamma_{\mathrm{cut}}} e^{iq_xx} \sqrt{q_x - k_0} \mathrm{d}q_x \right)
\end{equation}

Notice that the remaining integral inside the parentheses is exactly the branch cut contribution $I(x)$ we rigorously evaluated in Appendix~\ref{sec:asymptotic_expansion}. By directly utilizing the result Eq.~\eqref{I(x)}, we obtain:
\begin{equation}
    I \approx -\frac{r_0 \sqrt{2k_0}}{2\pi} \left[ \frac{\sqrt{\pi}}{x^{3/2}} e^{i(k_0 x + 3\pi/4)} \right]
\end{equation}

Simplifying the constant coefficients yields:
\begin{equation}
    I \approx - r_0 \sqrt{\frac{k_0}{2\pi}} \frac{1}{x^{3/2}} e^{i(k_0 x + 3\pi/4)}
\end{equation}
Thus, we finally arrive at the far-field current of Eq.~\eqref{j_yt}.

\section{Evaluation of the Branch Cut Integral}
\label{sec:asymptotic_expansion}

In the calculation of the far-field ($x \to \infty$) asymptotic expansion for both static and dynamic case, the primary contribution arises from the integral surrounding the branch cut. This section provides the detailed derivation of the core integral evaluated along the ``hairpin'' contour $\Gamma_{\mathrm{cut}}$. We consider the following integral form:
\begin{equation}
    I(x) = -I_{\mathrm{cut}} = -\int_{\Gamma_{\mathrm{cut}}} e^{iqx} \sqrt{q - k_0} \mathrm{d}q, \quad (x > 0)
    \label{D1}
\end{equation}
where $q$ is the complex variable, and $k_0$ represents the branch point. To satisfy the requirement of causality in physical systems, we introduce an infinitesimal positive imaginary part to the frequency, effectively shifting the branch point into the upper half-plane (UHP), i.e., $k_0 \to k_0 + i0^+$.

\subsection{Parameterization and Discontinuity Evaluation}
The branch cut is defined as a vertical line extending from the branch point $k_0$ to $k_0 + i\infty$ in the complex $q$-plane. We parameterize the integration path by setting $q = k_0 + is$, where $s \in (0, \infty)$ is a real variable representing the coordinate along the cut, and $\mathrm{d}q = i \mathrm{d}s$. 

The hairpin contour $\Gamma_{\mathrm{cut}}$ consists of two paths: a downward path along the right side of the cut and an upward path along the left side. The total contour integral $I_{\mathrm{cut}}$ can be expressed as:
\begin{equation}
    I_{\mathrm{cut}} = \int_{\infty}^{0} f_{\mathrm{right}}(s) i\mathrm{d}s + \int_{0}^{\infty} f_{\mathrm{left}}(s) i\mathrm{d}s = -i \int_{0}^{\infty} \left[ f_{\mathrm{right}}(s) - f_{\mathrm{left}}(s) \right] \mathrm{d}s,
\end{equation}
where $f_{\mathrm{right/left}}(s)$ denotes the value of the integrand $e^{iqx} \sqrt{q - k_0}$ on the respective side of the cut. Substituting $I(x) = -I_{\mathrm{cut}}$, we obtain:
\begin{equation}
    I(x) = i \int_{0}^{\infty} \left[ f_{\mathrm{right}}(s) - f_{\mathrm{left}}(s) \right] \mathrm{d}s.
    \label{eq:jump_integral}
\end{equation}

To evaluate the discontinuity of the multi-valued function $\sqrt{q-k_0}$, we utilize the exponential representation. For $q - k_0 = s e^{i\theta}$, the phase on the right side of the cut is $\theta = \pi/2$:
\begin{equation}
    \sqrt{(q-k_0)_{\mathrm{right}}} = \sqrt{s e^{i\pi/2}} = \sqrt{s} e^{i\pi/4}.
\end{equation}
As the contour wraps around the branch point $k_0$ to the left side, the phase advances by $2\pi$ to $\theta = 5\pi/2$. Crossing the Riemann sheet introduces a sign change:
\begin{equation}
    \sqrt{(q-k_0)_{\mathrm{left}}} = \sqrt{s e^{i5\pi/2}} = \sqrt{s} e^{i( \pi/4 + \pi)} = -\sqrt{s} e^{i\pi/4}.
\end{equation}
The term $e^{iqx}$ remains continuous across the cut and is given by $e^{i(k_0 + is)x} = e^{ik_0 x} e^{-sx}$. Thus, the discontinuity across the branch cut is:
\begin{equation}
    f_{\mathrm{right}}(s) - f_{\mathrm{left}}(s) = e^{ik_0 x} e^{-sx} \left[ \sqrt{s} e^{i\pi/4} - \left( -\sqrt{s} e^{i\pi/4} \right) \right] = 2\sqrt{s} e^{i\pi/4} e^{ik_0 x} e^{-sx}.
\end{equation}

\subsection{Analytical Form of the Final Integral}
Substituting the derived discontinuity into Eq.~\eqref{eq:jump_integral}, and factoring out the constants, we have:
\begin{equation}
    I(x) = i \left( 2 e^{i\pi/4} e^{ik_0 x} \right) \int_{0}^{\infty} \sqrt{s} e^{-sx} \mathrm{d}s.
\end{equation}
The integral over $s$ can be evaluated using the the Gamma function, $\int_{0}^{\infty} s^{n} e^{-sx} \mathrm{d}s = \Gamma(n+1)/x^{n+1}$. For $n=1/2$:
\begin{equation}
    \int_{0}^{\infty} s^{1/2} e^{-sx} \mathrm{d}s = \frac{\Gamma(3/2)}{x^{3/2}} = \frac{\sqrt{\pi}}{2x^{3/2}}.
\end{equation}
By applying Euler's formula $i = e^{i\pi/2}$, the final expression for the branch cut integral simplifies to:
\begin{equation}
    I(x) = e^{i\pi/2} \cdot 2 e^{i\pi/4} e^{ik_0 x} \cdot \frac{\sqrt{\pi}}{2x^{3/2}} = \frac{\sqrt{\pi}}{x^{3/2}} e^{i(k_0 x + 3\pi/4)}.\label{I(x)}
\end{equation}
This result characterizes the $|x|^{-3/2}$ power-law decay of the dynamic current in the far-field limit.

\section{Far-field expansion with physical pole contribution}

We now consider the long-range asymptotic behavior of $\boldsymbol{j}_x$:
\begin{equation}
    I(x) = \int_{-\infty}^{\infty} \frac{\mathrm{d}q_x}{2\pi} e^{iq_xx} \frac{r_0 \omega^2}{c^2 \sqrt{q_x^2 - k_0^2} - r_0 \omega^2}, \quad (x > 0)
\end{equation}
To simplify the notation, we introduce $\alpha = r_0 \omega^2 / c^2 > 0$. The integral becomes:
\begin{equation}
    I(x) = \int_{-\infty}^{\infty} \frac{\mathrm{d}q_x}{2\pi} e^{iq_xx} \frac{\alpha}{\sqrt{q_x^2 - k_0^2} - \alpha},
\end{equation}

The condition for the pole in this integrand is $\sqrt{q_x^2 - k_0^2} = \alpha > 0$. We define:
\begin{equation}
    q_p = \sqrt{k_0^2 + \alpha^2}.
\end{equation}

The equation gives two poles $\pm q_p$. Taking causality into account ($\omega \to \omega + i0^+$), the positive pole shifts into the upper half-plane ($q_p \to q_p + i0^+$), while the negative pole shifts into the lower half-plane. Since we close the integration contour in the upper half-plane, only the pole at $+q_p$ contributes to the integral.

Since the integrand vanishes as $|q_x| \to \infty$, we can safely deform the integration contour into the UHP, encompassing both the branch cut originating at $k_0$ and the physical pole at $q_p$. According to Cauchy's residue theorem, the integral is given by:
\begin{equation}
    I_{\mathrm{real}} = 2\pi i \cdot \mathrm{Res}(q_p) - I_{\mathrm{cut}}
    \label{eq:contour_pole},
\end{equation}
where $I_{\mathrm{cut}}$ is the branch cut integral evaluated along the hairpin contour $\Gamma_{\mathrm{cut}}$.

Let  $D(q_x) = \sqrt{q_x^2 - k_0^2} - \alpha$. The derivative evaluated at the pole is $D'(q_p) = q_p / \alpha$. The residue of the integrand $g(q_x) = \frac{1}{2\pi} e^{iq_xx} \frac{\alpha}{D(q_x)}$ at $q_x = q_p$ is:
\begin{equation}
    \mathrm{Res}(g, q_p) = \frac{\frac{1}{2\pi} e^{iq_p x} \alpha}{D'(q_p)} = \frac{\alpha^2}{2\pi q_p} e^{iq_p x}.
\end{equation}
Thus, the pole contribution to the integral is:
\begin{equation}
    2\pi i \cdot \mathrm{Res}(g, q_p) = i \frac{\alpha^2}{q_p} e^{iq_p x}.
    \label{eq:pole_result}
\end{equation}

For the branch cut integration, the exponential decay along $\Gamma_{\mathrm{cut}}$ confines the dominant contribution to the vicinity of the branch point $q_x \approx k_0$ in the far-field limit. Within this local region, we can safely Taylor expand the fraction:
\begin{equation}
    \frac{\alpha}{\sqrt{q_x^2 - k_0^2} - \alpha} = -1 - \frac{\sqrt{q_x^2 - k_0^2}}{\alpha} + \mathcal{O}(q_x^2 - k_0^2).
\end{equation}
The constant term $-1$ is continuous across the cut and therefore contributes nothing to the contour integral around it. Thus, the long-range asymptotic behavior of the cut contribution is dominated by the leading term $-\frac{1}{\alpha}\sqrt{q_x^2 - k_0^2}$ evaluated along $\Gamma_{\mathrm{cut}}$. By approximating $\sqrt{q_x^2 - k_0^2} \approx \sqrt{2k_0}\sqrt{q_x - k_0}$, we notice that this maps exactly to the branch cut integral evaluated in Section~\ref{sec:asymptotic_expansion}. The branch cut contribution $-I_{\mathrm{cut}}$ to the total integral evaluates to:
\begin{equation}
    -I_{\mathrm{cut}} \approx -\frac{1}{2\pi\alpha} \left( \sqrt{2\pi k_0} \frac{1}{x^{3/2}} e^{i(k_0 x + 3\pi/4)} \right) = -\frac{1}{\alpha} \sqrt{\frac{k_0}{2\pi}} \frac{1}{x^{3/2}} e^{i(k_0 x + 3\pi/4)},
    \label{eq:cut_result}
\end{equation}

Summing the components from Eq.~\eqref{eq:pole_result} and Eq.~\eqref{eq:cut_result}, the complete asymptotic expansion in the far-field is:
\begin{equation}
    I(x) \sim i \frac{\alpha^2}{\sqrt{k_0^2 + \alpha^2}} e^{i \sqrt{k_0^2 + \alpha^2} x} - \frac{1}{\alpha} \sqrt{\frac{k_0}{2\pi}} \frac{1}{x^{3/2}} e^{i(k_0 x + 3\pi/4)},
\end{equation}

\section{Magnetic Impurity Case}
\label{sec:time_dependent}

To further investigate superconductor-mediated magnetic interactions, we consider a zero-dimensional magnetic impurity. The impurity is located at the origin of an infinite two-dimensional superconducting film.

The full gauge-invariant supercurrent density $\boldsymbol{j}$ in the presence of an electromagnetic field $\boldsymbol{A}$ and an effective gauge field $\boldsymbol{a}$ is given by:
\begin{equation}
    \boldsymbol{j}(\boldsymbol{r}, t) = -e^2 D \left[ \boldsymbol{A}(\boldsymbol{r}, t) +  \boldsymbol{a}(\boldsymbol{r}, t) - \frac{1}{2e} \nabla \phi(\boldsymbol{r}, t) \right].
    \label{eq:current_def}
\end{equation}

According to the Helmholtz decomposition, vector fields can be separated into transverse ($\nabla \cdot \boldsymbol{v}_t = 0$) and longitudinal ($\nabla \times \boldsymbol{v}_l = 0$) components. In the Lorenz gauge ($\nabla \cdot \boldsymbol{A} + c^{-2}\partial_t \Phi = 0$), the wave equation decouples into independent transverse and longitudinal sectors. Since the scalar phase gradient $\nabla \phi$ is purely longitudinal, it does not contribute to the transverse supercurrent. Focusing on the transverse radiative modes, the transverse supercurrent density $\boldsymbol{j}_t$ is written as:
\begin{equation}
    \boldsymbol{j}_t(\boldsymbol{r}, t) = -e^2 D \left[ \boldsymbol{A}_t(\boldsymbol{r}, t) + \boldsymbol{a}_t(\boldsymbol{r}, t) \right].
    \label{eq:current_def_transverse}
\end{equation}

The transverse vector potential $\boldsymbol{A}_t$ is driven by $\boldsymbol{j}_t$ through the wave equation:
\begin{equation}
    \left( \nabla^2 - \frac{1}{c^2} \frac{\partial^2}{\partial t^2} \right) \boldsymbol{A}_t(\boldsymbol{r}, z, t) = -\mu_0 \boldsymbol{j}_t(\boldsymbol{r}, t)\delta(z).
    \label{eq:wave_eq}
\end{equation}

Assuming a harmonic time dependence $e^{-i\omega t}$, Eq.~\eqref{eq:wave_eq} becomes a Helmholtz equation. Performing a Fourier transform with respect to the in-plane coordinates $\boldsymbol{r}=(x,y)$ with momentum $\boldsymbol{q}$, and substituting $\boldsymbol{j}_t$ from Eq.~\eqref{eq:current_def_transverse}, the equation for the Fourier component $\boldsymbol{A}_t(\boldsymbol{q}, z)$ reads:
\begin{equation}
    \left( \frac{\partial^2}{\partial z^2} - \kappa^2\right) \boldsymbol{A}_t(\boldsymbol{q}, z)  
    = \frac{2}{r_0} \left[ \boldsymbol{A}_t(\boldsymbol{q}, 0) + \boldsymbol{a}_t(\boldsymbol{q}) \right] \delta(z),
    \label{eq:fourier_maxw}
\end{equation}
where $\kappa = \sqrt{q^2 - k_0^2}$, $q^2=q_x^2+q_y^2$, $k_0 = \omega/c$, and $r_0 = 2/(\mu_0 e^2 D)$ is the Pearl length. The solution decaying away from the film plane ($|z|>0$) is $\boldsymbol{A}_t(\boldsymbol{q}, z) = \boldsymbol{A}_t(\boldsymbol{q}, 0)e^{-\kappa |z|}$. 

Integrating Eq.~\eqref{eq:fourier_maxw} across the $z=0$ plane yields the boundary condition:
\begin{equation}
    \frac{\partial \boldsymbol{A}_t}{\partial z}\bigg|_{0^+} - \frac{\partial \boldsymbol{A}_t}{\partial z}\bigg|_{0^-} = \frac{2}{r_0} \left[ \boldsymbol{A}_t(\boldsymbol{q}, 0) + \boldsymbol{a}_t(\boldsymbol{q}) \right].
\end{equation}

This gives $-2\kappa \boldsymbol{A}_t(\boldsymbol{q}, 0) = (2/r_0) [\boldsymbol{A}_t(\boldsymbol{q}, 0) + \boldsymbol{a}_t(\boldsymbol{q})]$, which leads to the solution for the transverse vector potential in the superconductor film:
\begin{equation}
    \boldsymbol{A}_t(\boldsymbol{q}, 0) = -\frac{1}{1 + \kappa r_0} \boldsymbol{a}_t(\boldsymbol{q}).
\end{equation}

Substituting this back into Eq.~\eqref{eq:current_def_transverse}, we obtain the transverse supercurrent in Fourier space:
\begin{equation}
    \boldsymbol{j}_t(\boldsymbol{q}) = -e^2 D \left( \frac{\kappa r_0}{1 + \kappa r_0} \right) \boldsymbol{a}_t(\boldsymbol{q}),
    \label{eq:current_q}
\end{equation}

For completeness, the total supercurrent $\boldsymbol{j}(\boldsymbol{q}, \omega)$ incorporates both the transverse radiative response and the longitudinal response. By decoupling the dynamic equations in the Lorenz gauge, the full current evaluates to:
\begin{equation}
    \boldsymbol{j}(\boldsymbol{q}, \omega) = -e^2 D \left[ \frac{\kappa r_0}{1 + \kappa r_0} \right] \boldsymbol{a}_t(\boldsymbol{q}, \omega) + e^2 D \left[ \frac{r_0 \omega^2}{c^2 \kappa - r_0 \omega^2} \right] \boldsymbol{a}_l(\boldsymbol{q}, \omega),
    \label{eq:current_full_q}
\end{equation}
where $\boldsymbol{a}_l(\boldsymbol{q})$ is the longitudinal component of the effective gauge field. Since our primary goal is to investigate long-range magnetic coupling mediated by transverse electromagnetic radiation, we proceed by evaluating transverse current.

Transforming back to real space, the current distribution is given by the convolution of the transverse effective gauge field with a frequency-dependent kernel:
\begin{equation}
    \boldsymbol{j}_t(\boldsymbol{r},\omega) = -e^2 D \bigg[ \boldsymbol{a}_t(\boldsymbol{r}) - \int d^2 r' \, K_\omega(|\boldsymbol{r} - \boldsymbol{r}'|) \boldsymbol{a}_t(\boldsymbol{r}') \bigg],
    \label{eq:current_real_space}
\end{equation}
where the kernel $K_\omega(R)$ is:
\begin{equation}
    K_\omega(R) = \int_0^\infty \frac{dq}{2\pi} \frac{q J_0(qR)}{1 + r_0 \sqrt{q^2 - (\omega/c)^2}}.
\end{equation}
The derivation details of this kernel are provided in Appendix \ref{K_omega}. $J_0(x)$ is the zeroth-order Bessel function. 

Using polar coordinates $(r, \theta)$ with orthogonal unit vectors $\hat{r}$ and $\hat{\theta}$, the planar source vector $\boldsymbol{a}$ is projected into radial and azimuthal components as $a_r = \boldsymbol{a} \cdot \hat{r}$ and $a_\theta = \boldsymbol{a} \cdot \hat{\theta}$. The resulting transverse current components are:
\begin{equation}
    \boldsymbol{j}_t(r,\theta,t)=\left[j_r(r,\theta)\hat{r}+j_{\theta}(r,\theta)\hat{\theta}\right]e^{-i\omega t},
\end{equation}
with:
\begin{subequations}
\begin{align}
    j_r(r, \theta) &= -e^2D a_r \int_0^\infty \frac{q \, dq}{2\pi} \left[ \frac{r_0\sqrt{q^2-k_0^2}}{1+r_0\sqrt{q^2-k_0^2}} \right] \frac{J_1(qr)}{qr}, \\
    j_\theta(r, \theta) &= -e^2D a_{\theta} \int_0^\infty \frac{q \, dq}{2\pi} \left[ \frac{r_0\sqrt{q^2-k_0^2}}{1+r_0\sqrt{q^2-k_0^2}} \right] J_1'(qr),
\end{align}
\end{subequations}
where $J_1(x)$ is the first-order Bessel function, and $J_1'(x) = dJ_1(x)/dx$.

\subsection{Far-Field Approximation}
\label{subsec:far_field}

We evaluate the asymptotic behavior of the current components in the far-field region ($k_0 r \gg 1$):
\begin{subequations}
\label{eq:asymptotic_currents}
\begin{align}
    j_r(r,\theta) &\approx -e^2 D a_r \left( \frac{-i k_0 r_0}{1 - i k_0 r_0} \right) \frac{1}{2\pi r^2} \label{eq:j_r_dyn}, \\
    j_\theta(r,\theta) &\approx e^2 D a_\theta \left( \frac{-i k_0 r_0}{1 - i k_0 r_0} \right) \frac{1}{2\pi r^2} 
    - i \frac{e^2 D a_\theta k_0 r_0}{2\pi r^2} e^{i k_0 r}.\label{eq:j_theta_dyn}
\end{align}
\end{subequations}

In the static limit ($\omega \to 0$), the dynamic radiative terms vanish. The static current $\boldsymbol{j}^{\text{stat}}$ can be evaluated using the limit $K(q) \approx r_0 q$, yielding a faster decay:
\begin{equation}
    j_r^{\text{stat}}(r) \approx -\frac{e^2 D a_r r_0}{2\pi r^3}, \quad j_\theta^{\text{stat}}(r) \approx \frac{e^2 D a_\theta r_0}{\pi r^3}.
\end{equation}

This derivation recovers the static limit, consistent with previous theoretical results for discrete magnetic impurities on spin-orbit-coupled superconductors \cite{lu2023ferromagnetic}. In the zero-frequency case, the supercurrent decays as $1/r^3$.

\subsection{Different Topology Case}

To explore the effects of different boundary topologies, we transition from a planar film to a spherical geometry. Assuming an isotropic radial Rashba spin-orbit coupling, we consider a localized magnetic impurity at the coordinates $(R, \theta_0, \varphi_0)$, where $\hat{r}_0$ denotes the unit radial vector pointing to the impurity.

Similar to the planar case, we focus on the transverse modes responsible for long-range coupling. On a spherical surface, any strictly transverse vector field can be generated by the surface curl operator. Thus, we express the transverse effective gauge field $\boldsymbol{a}_t(\boldsymbol{r})$ and the transverse supercurrent density $\boldsymbol{j}_t(\boldsymbol{r})$ as:
\begin{subequations}
\begin{align}
    \boldsymbol{a}_t(\boldsymbol{r}) &= -\hat{r} \times \nabla_{||} \alpha(\theta, \varphi),\\
    \boldsymbol{j}_t(\boldsymbol{r}) &= -\hat{r} \times \nabla_{||} \psi(\theta, \varphi),
\end{align}
\end{subequations}
where the surface gradient operator is defined as:
\begin{equation}
    \nabla_{||} = \hat{\theta} \frac{1}{R} \frac{\partial}{\partial \theta} + \hat{\varphi} \frac{1}{R \sin\theta} \frac{\partial}{\partial \varphi}.
\end{equation}

The scalar potentials are expanded in terms of spherical harmonics:
\begin{subequations}
\begin{align}
    \alpha(\theta, \varphi) &= \sum_{l=1}^{\infty} \sum_{m=-l}^{l} a_{lm} Y_{lm}(\theta, \varphi), \\
    \psi(\theta, \varphi) &= \sum_{l=1}^{\infty} \sum_{m=-l}^{l} \psi_{lm} Y_{lm}(\theta, \varphi).
\end{align}
\end{subequations}
For convenience, we define the transverse vector multipole modes as $\boldsymbol{a}_{t,lm}(\theta, \varphi) = a_{lm} \left[ -\hat{r} \times \nabla_{||} Y_{lm}(\theta,\varphi) \right]$ and $\boldsymbol{j}_{t,lm}(\theta, \varphi) = \psi_{lm} \left[ -\hat{r} \times \nabla_{||} Y_{lm}(\theta,\varphi) \right]$.

The spatial distribution of the localized source is parameterized by a delta function:
\begin{equation}\label{eq:a_r_r}
    \boldsymbol{a}_t(\boldsymbol{r}) = \boldsymbol{W}_t \frac{\delta(\theta - \theta_0)\delta(\varphi - \varphi_0)}{R^2 \sin\theta},
\end{equation}
where $\boldsymbol{W}_t$ is the transverse component of the impurity magnetization ($\boldsymbol{W}_t \cdot \hat{r}_0 = 0$). By projecting Eq.~\eqref{eq:a_r_r} onto the vector spherical harmonic basis and integrating over the solid angle, we extract the expansion coefficients:
\begin{equation}
    a_{lm} = -\frac{1}{l(l+1)} \boldsymbol{W}_t \cdot \left[ \hat{r}_0 \times \nabla_{||}Y_{lm}^*(\theta_0, \varphi_0) \right].
\end{equation}

To determine the induced supercurrent, we employ the transverse constitutive relation in the multipole basis:
\begin{equation}\label{eq:j_constitutive}
    \boldsymbol{j}_{t,lm} = -e^2D(\boldsymbol{A}_{t,lm} + \boldsymbol{a}_{t,lm}).
\end{equation}
For a current confined to a spherical surface of radius $R$, classical electrodynamics dictates that the magnetic vector potential for the $l$-th multipole is proportional to the current mode, given by $\boldsymbol{A}_{t,lm} = \frac{\mu_0 R}{2l+1} \boldsymbol{j}_{t,lm}$. Substituting this into Eq.~\eqref{eq:j_constitutive} yields:
\begin{equation}
    \boldsymbol{j}_{t,lm} = -e^2D \left( \frac{\mu_0 R}{2l+1} \boldsymbol{j}_{t,lm} + \boldsymbol{a}_{t,lm} \right).
\end{equation}
Solving for the current modes, we obtain the coefficient relation:
\begin{equation}
    \psi_{lm} = -e^2D \left[ \frac{(2l+1)r_0}{(2l+1)r_0 + 2R} \right] a_{lm},
\end{equation}
where $r_0 = 2/(\mu_0 e^2 D)$ is the Pearl length.

Furthermore, we extend our spherical framework into the dynamic regime. Assuming the transverse effective gauge field oscillates with a frequency $\omega$, it takes the form $\boldsymbol{a}_t(\boldsymbol{r}, t) = \boldsymbol{a}_t(\boldsymbol{r}) e^{-i\omega t}$. We introduce the wave number $k_0 = \omega/c$.

In the dynamic case, capturing electromagnetic retardation requires the full wave equation. Expanding the retarded Green's function in terms of spherical Bessel ($j_l(k_0 R)$) and Hankel ($h_l^{(1)}(k_0 R)$) functions, the electrodynamic feedback for the $l$-th multipole becomes complex and frequency-dependent:
\begin{equation}
    \boldsymbol{A}_{t,lm}(\omega) = \left[ \mu_0 R^2 i k_0 j_l(k_0 R) h_l^{(1)}(k_0 R) \right] \boldsymbol{j}_{t,lm}(\omega).
\end{equation}
Here, the spherical Bessel function $j_l$ accounts for the standing wave modes on the sphere, while the Hankel function $h_l^{(1)}$ describes the outward-propagating radiative waves. 

Substituting this dynamic vector potential into the transverse constitutive relation $\boldsymbol{j}_{t,lm}(\omega) = -e^2D[\boldsymbol{A}_{t,lm}(\omega) + \boldsymbol{a}_{t,lm}(\omega)]$ and solving for the current modes, we obtain the dynamic coefficient:
\begin{equation}
    \psi_{lm}(\omega) = -e^2D \left[ \frac{1}{1 + i \frac{2 k_0 R^2}{r_0} j_l(k_0 R) h_l^{(1)}(k_0 R)} \right] a_{lm}.
\end{equation}

By summing over all multipole modes, we arrive at the expression for the time-dependent transverse supercurrent distribution on the sphere:
\begin{align} 
    \boldsymbol{j}_t(\theta, \varphi, t) = e^2 D \sum_{l=1}^{\infty} \sum_{m=-l}^{l} \left[ \frac{1}{1 + i \frac{2 k_0 R^2}{r_0} j_l(k_0 R) h_l^{(1)}(k_0 R)} \right] \nonumber 
     a_{lm} \left(\hat{r} \times \nabla_{||} Y_{lm}(\theta, \varphi) \right) e^{-i\omega t}.\label{eq:dynamic_spherical_current}
\end{align}

In the zero-frequency limit ($\omega \rightarrow 0$), the kernel recovers the factor $r_0(2l+1) / [r_0(2l+1) + 2R]$, consistent with the static case.

\section{Derivation of the kernel \texorpdfstring{$K_\omega(R)$}{K\_omega(R)}}
\label{K_omega}

In Section~\ref{sec:time_dependent}, we introduced the current generated by magnetic impurities on the surface of infinite superconductor thin film see in Eq.~\eqref{eq:current_real_space}. Now we derive the kernel $K_\omega(R)$ in this section. We need to calculate the Inverse Fourier transform of the term as follows:
\begin{equation}
    F^{-1} \left\{ \left( \frac{1}{1 + r_0 \kappa(q)} \right) a_t(q) \right\},
\end{equation}
where $\kappa(q)=\sqrt{q^2-(\omega/c)^2}$, $q=\sqrt{q_x^2+q_y^2}$. According to the Convolution Theorem of Fourier Transform, we can write:
\begin{equation}
    {F}^{-1} \{ K(q) a_t(q) \} = K(\boldsymbol{r}) * a_t(\boldsymbol{r}) = \int d^2 r' K(\boldsymbol{r} - \boldsymbol{r}') a_t(\boldsymbol{r}'),
\end{equation}
Let $\boldsymbol{R} = \boldsymbol{r} - \boldsymbol{r}'$, and $R = |\boldsymbol{R}|$, we get:
\begin{equation}
    K_\omega(\boldsymbol{R}) = \int \frac{d^2 q}{(2\pi)^2} \frac{1}{1 + r_0 \sqrt{q^2 - (\omega/c)^2}} e^{i \boldsymbol{q} \cdot \boldsymbol{R}}.
\end{equation}
Since the denominator of the integral kernel depends only on the magnitude of the momentum $q = |\boldsymbol{q}|$, we introduce two-dimensional polar coordinates. Let the polar angle of vector $\boldsymbol{q}$ be $\theta_q$, and the polar angle of the space vector $\boldsymbol{R}$ be $\theta_R$. Therefore, $\boldsymbol{q} \cdot \boldsymbol{R} = q R \cos(\theta_q - \theta_R)$ and
$d^2 q = q \, dq \, d\theta_q$. The kernel becomes:
\begin{equation}
    K_\omega(R) = \int_0^\infty \frac{q \, dq}{(2\pi)^2} \frac{1}{1 + r_0 \sqrt{q^2 - (\omega/c)^2}} \int_0^{2\pi} d\theta_q \, e^{i q R \cos(\theta_q - \theta_R)}.
\end{equation}
The angular integral is equal to the zeroth-order Bessel function of the first kind, $J_0(x)$, multiplied by $2\pi$.
\begin{equation}
    \int_0^{2\pi} d\theta_q \, e^{i q R \cos(\theta_q - \theta_R)} = 2\pi J_0(qR),
\end{equation}
so we finally obtain the kernel:
\begin{equation}
    K_\omega(R) = \int_0^\infty \frac{dq}{2\pi} \frac{q J_0(qR)}{1 + r_0 \sqrt{q^2 - (\omega/c)^2}}.
\end{equation}

\section{Derivation of the current in magnetic impurity case}
According to Eq.~\eqref{eq:current_q}:
$\boldsymbol{a}_t(\boldsymbol{q})$ is the transverse component of $\boldsymbol{a}$. We can write as:
\begin{equation}
    \boldsymbol{a}_t(\boldsymbol{q}) = \boldsymbol{a} - \frac{\boldsymbol{q} (\boldsymbol{q} \cdot \boldsymbol{a})}{q^2},\label{a_t}
\end{equation}
For the sake of simplicity, let $f_t(q) =  \left( \frac{r_0 \kappa(q)}{1 + r_0 \kappa(q)} \right)$.
We take polar coordinates $(r, \theta)$ in real space. At the point $\boldsymbol{r}$, we define a radial unit vector $\hat{r}$ and an angular unit vector $\hat{\theta}$. Since $\boldsymbol{a}$ is a constant vector, we can project it onto this local polar coordinate basis:
\begin{equation}
    \boldsymbol{a} = a_r \hat{r} + a_\theta \hat{\theta}.
\end{equation}
The current is given by:
\begin{equation}
    \boldsymbol{j}(\boldsymbol{r}) = \int \frac{d^2 q}{(2\pi)^2} e^{i \boldsymbol{q} \cdot \boldsymbol{r}} \boldsymbol{j}(\boldsymbol{q}).\label{j_(r)real}
\end{equation}

When performing the $d^2 q$ integration, the wave vector $\boldsymbol{q}$ sweeps across all directions. Let $\beta$ be the angle between $\boldsymbol{q}$ and  $\boldsymbol{r}$.
Then, the wave vector $\boldsymbol{q}$ can also be written in this basis: $\boldsymbol{q} = q (\cos\beta \, \hat{r} + \sin\beta \, \hat{\theta})$, and we also have $e^{i \boldsymbol{q} \cdot \boldsymbol{r}} = e^{i q r \cos\beta}$ . We substitute $\boldsymbol{a}$ and $\boldsymbol{q}$ into the Eq.~\eqref{a_t}:
\begin{align}
    \boldsymbol{a}_t(\boldsymbol{q}) &=  \boldsymbol{a} - \frac{\boldsymbol{q} (\boldsymbol{q} \cdot \boldsymbol{a})}{q^2}\nonumber\\
    &= a_r\hat{r}+a_\theta\hat{\theta} - (\cos\beta \, \hat{r} + \sin\beta \, \hat{\theta}) (a_r \cos\beta + a_\theta \sin\beta)\nonumber\\
    &= [a_r - (a_r \cos^2\beta + a_\theta \sin\beta \cos\beta)] \hat{r} + [a_\theta - (a_r \sin\beta \cos\beta + a_\theta \sin^2\beta)] \hat{\theta}\nonumber\\
    &=[a_r \sin^2\beta - a_\theta \sin\beta \cos\beta] \hat{r} + [-a_r \sin\beta \cos\beta + a_\theta \cos^2\beta] \hat{\theta}.\label{a_t(q)}
\end{align}
According to the definition of Bessel functions, we can write the following integrals:
\begin{align}
    I_1 &= \int_0^{2\pi} e^{i q r \cos\beta} \sin^2\beta \, d\beta= 2\pi \frac{J_1(qr)}{qr}\\
    I_2 &= \int_0^{2\pi} e^{i q r \cos\beta} \sin\beta \cos\beta \, d\beta=0\\
    I_3 &= \int_0^{2\pi} e^{i q r \cos\beta} \cos^2\beta \, d\beta= 2\pi J_1'(qr).
\end{align}
Substitute Eq.~\eqref{a_t(q)} into Eq.~\eqref{j_(r)real}:
\begin{align}
    j_r &=-e^2 D \int_0^\infty \frac{q \, dq}{(2\pi)^2} f_t(q) \left[ a_r I_1 - a_\theta I_2 \right]\\
    j_\theta &=-e^2 D \int_0^\infty \frac{q \, dq}{(2\pi)^2} f_t(q) \left[ -a_r I_2 + a_\theta I_3 \right].
\end{align}
Finally, we get the current induced by the magnetic impurity in the infinity superconductor film:
\begin{align}
    j_r(r, \theta) &= -e^2D a_r \int_0^\infty \frac{q \, dq}{2\pi} \left[ \frac{r_0\sqrt{q^2-k_0^2}}{1+r_0\sqrt{q^2-k_0^2}} \right] \frac{J_1(qr)}{qr}, \\
    j_\theta(r, \theta) &= -e^2D a_{\theta} \int_0^\infty \frac{q \, dq}{2\pi} \left[ \frac{r_0\sqrt{q^2-k_0^2}}{1+r_0\sqrt{q^2-k_0^2}} \right] J_1'(qr).
\end{align}

\section{Far-field Asymptotic Expansion of \texorpdfstring{$j_r, j_\theta$}{j\_r, j\_theta}}
\label{Far-field Asymptotic Expansion}
In this section, we provide the derivation of the far-field asymptotic expansion ($r \to \infty$) for the radial and azimuthal currents. The integrals are governed by the kernel function:
\begin{equation}
    f_t(q) = \frac{r_0 \kappa(q)}{1 + r_0 \kappa(q)},
\end{equation}
where $\kappa(q) = \sqrt{q^2 - k_0^2}$. To satisfy causality, we apply  $\omega \to \omega + i0^+$, which ensures $k_0 \to k_0 + i0^+$. Consequently, at the $q=0$ limit, we have $\kappa(0) = -ik_0$. The zero-momentum limit of the kernel evaluates to:
\begin{equation}
    f_t(0) = \frac{-i k_0 r_0}{1 - i k_0 r_0}.
\end{equation}

The integral expression for the radial current is:
\begin{equation}
    j_r(r,\theta) = -e^2 D a_r \int_0^\infty \frac{q \, dq}{2\pi} f_t(q) \frac{J_1(qr)}{qr} = -e^2 D a_r \frac{1}{2\pi r} \int_0^\infty f_t(q) J_1(qr) \, dq.
\end{equation}
In the far-field limit ($r \to \infty$), the Bessel function $J_1(qr)$ oscillates rapidly. The dominant contribution to the integral arises from the $q \to 0$ region. By approximating $f_t(q) \approx f_t(0)$ and utilizing the standard Bessel integral identity $\int_0^\infty J_1(qr) \, dq = \frac{1}{r}$, we obtain the asymptotic radial current:
\begin{equation}
    j_r(r,\theta) \approx -e^2 D a_r f_t(0) \frac{1}{2\pi r} \left( \frac{1}{r} \right) = -e^2 D a_r \left( \frac{-i k_0 r_0}{1 - i k_0 r_0} \right) \frac{1}{2\pi r^2}.
\end{equation}

For the azimuthal current, the exact expression incorporates the derivative of the Bessel function:
\begin{equation}
    j_\theta(r,\theta) = -e^2 D a_\theta \int_0^\infty \frac{q \, dq}{2\pi} f_t(q) J_1'(qr).
\end{equation}
Applying the Bessel function recurrence relation $J_1'(qr) = J_0(qr) - \frac{1}{qr} J_1(qr)$, the integral is separated into two components:
\begin{equation}
    j_\theta(r,\theta) = e^2 D a_\theta \left[ \frac{1}{2\pi r} \int_0^\infty f_t(q) J_1(qr) \, dq - \frac{1}{2\pi} \int_0^\infty f_t(q) q J_0(qr) \, dq \right]=I_{sta}+I_{rad}.
\end{equation}
The first term is:
\begin{equation}
    I_{\text{sta}} = e^2 D a_\theta f_t(0) \frac{1}{2\pi r^2}.
\end{equation}
The second term is the dynamic radiative contribution. In the far-field, the integral $\int_0^\infty f_t(q) q J_0(qr) \, dq$ is dominated by the branch point $q = k_0$. Near this point, the kernel can be approximated by its leading term as $f_t(q) \approx r_0 \sqrt{q^2 - k_0^2}$. The integral becomes:
\begin{equation}
    I_{\text{rad}} \approx - \frac{e^2 D a_\theta r_0}{2\pi} \int_0^\infty q \sqrt{q^2 - k_0^2} J_0(qr) \, dq.\label{I_rad}
\end{equation}
To evaluate this integral, we use the Sommerfeld identity:
\begin{equation}
    \int_0^\infty \frac{q}{\sqrt{q^2 - k_0^2}} J_0(qr) e^{-z \sqrt{q^2 - k_0^2}} \, dq = \frac{e^{ik_0 \sqrt{r^2 + z^2}}}{\sqrt{r^2 + z^2}}.
\end{equation}
By applying the second partial derivative with respect to $z$, to both sides and setting $z=0$, the integrand on the left side is multiplied by $(-\sqrt{q^2 - k_0^2})^2 = q^2 - k_0^2$. The corresponding evaluation on the right side yields:
\begin{equation}
    \left. \frac{\partial^2}{\partial z^2} \left( \frac{e^{ik_0 \sqrt{r^2+z^2}}}{\sqrt{r^2+z^2}} \right) \right|_{z=0} = \frac{1}{r} \frac{\partial}{\partial r} \left( \frac{e^{ik_0 r}}{r} \right) = \frac{i k_0}{r^2} e^{ik_0 r} - \frac{1}{r^3} e^{ik_0 r}.
\end{equation}
Retaining the leading order $1/r^2$ term for the far-field expansion, the integral simplifies to $i k_0 e^{ik_0 r} / r^2$. Substituting this result back into Eq.~\eqref{I_rad} gives:
\begin{equation}
    I_{\text{rad}} = -i \frac{e^2 D a_\theta k_0 r_0}{2\pi r^2} e^{ik_0 r}.
\end{equation}
Summing the two terms $I_{sta},I_{rad}$, the complete asymptotic expansion for the azimuthal current is obtained as:
\begin{equation}
    j_\theta(r,\theta) \approx e^2 D a_\theta \left[ \frac{f_t(0)}{2\pi r^2} - i \frac{k_0 r_0}{2\pi r^2} e^{ik_0 r} \right].
\end{equation}

\end{document}